\documentclass[prb,a4paper,showpacs,twocolumn,superscriptaddress,longbibliography]{revtex4-1}

\usepackage{textcomp}
\usepackage{amssymb}
\usepackage{amsmath}
\usepackage{amsfonts}
\usepackage{graphicx}
\usepackage{bm}
\usepackage{wasysym}
\usepackage{xcolor}
\usepackage{multirow}
\usepackage{natbib}
\usepackage{hyperref}
\usepackage{mathrsfs}

\begin{document}

\title{Weyl point immersed in a continuous spectrum: an example from superconducting nanostructures}

\author{Y. Chen}

\affiliation{Kavli Institute of Nanoscience, Delft University of Technology, 2628 CJ Delft, The Netherlands}

\author{Y. V. Nazarov}

\affiliation{Kavli Institute of Nanoscience, Delft University of Technology, 2628 CJ Delft, The Netherlands}

\begin{abstract}
A Weyl point in a superconducting nanostructure is a generic minimum model of a topological singularity at low energies. We connect the nanostructure to normal leads thereby immersing the topological singularity in the continuous spectrum of the electron states in the leads. This sets another simple and generic model useful to comprehend the modification of low-energy signularity in the presence of continuous spectrum. 
The tunnel coupling to the leads gives rise to new low energy scale $\Gamma$ at which all topological features are smoothed. We investigate superconducting and normal currents in the nanostructure at this scale. We show how the tunnel currents can be used for detection of the Weyl point. Importantly, we find that the topological charge is not concentrated in a point but rather is spread over the parameter space in the vicinity of the point. We introduce and compute the resulting topological charge density. We also reveal that the pumping to the normal leads helps to detect and investigate the topological effects in the vicinity of the point.  
\end{abstract}

\maketitle

\section{Introduction}
\label{sec:intro}
The study of topological materials has been on the front edge of the modern research in condensed matter physics for the past decade \cite{PhysRevLett.121.087001, PhysRevLett.121.086803, PhysRevLett.121.037701, PhysRevLett.120.256601, PhysRevLett.120.130503}. These materials are appealing from fundamental point of view and for possible applications \cite{PhysRevB.97.081402, PhysRevApplied.8.064001}, \cite{PhysRevApplied.2.054010}, \cite{PhysRevB.82.195409},\cite{PhysRevLett.107.056804}, including quantum information processing\cite{RevModPhys.80.1083, PhysRevX.6.031016}). The basis for applications is the topological protection of quantum states, which makes the states robust against small perturbations and leads to many unusual phenomena, e.g. topologically protected edge states\cite{PhysRevLett.95.226801, PhysRevLett.96.106401, PhysRevLett.98.106803}. The topological superconductors\cite{PhysRevLett.102.187001, PhysRevB.82.184516, nphys2479, PhysRevLett.105.097001} and Chern insulators\cite{PhysRevLett.61.2015, PhysRevX.1.021014, PhysRevB.89.195144, PhysRevB.74.235111} are the classes of topological materials that are under active investigation.

Most topological effects under consideration require discrete quantum states, for instance, electron, photon or phonon bands in a Brillouin zone of a periodic structure. Topological protection requires a gap in energy spectrum, that is, absence of continuous excitation spectrum at low energies.  It is intuitively clear that immersing the discrete states in a continuous spectrum, and compromising the energy gaps in this way will lead to compromising the topology. One of the goals of the present paper to propose and investigate a simple model for this that can be elaborated analytically to all details.

We concentrate on Weyl points those are most generally defined as topologically protected crossings of the discrete energy levels in a parametric space. From general topological reasoning, such crossing requires tuning of three parameters, so it is natural to consider a three-dimensional parametric space.


Recently, Weyl points - the topologically protected crossings in the spectrum of Andreev bound states  - have been predicted in superconducting nanostructures\cite{Weyl}. The specifics of superconductivity that these crossings may be pinned to Fermi level. This restricts the relevant physics to low energies and the properties of the ground state of the system.  At a Weyl point, the energy of the lowest Andreev state crosses Fermi level, so it costs vanishing energy to excite a quasiparticle in the vicinity of the point. A requirement of realization. This is why the Weyl points are usually considered in multi-terminal superconducting nanostructures where the parameters are the superconducting phase differences of the terminals. Four terminals are thus needed to realize a Weyl point. This prediction gave rise to related experimental and theoretical research \cite{Manucharyan2020, Belzig2020, Pribiag2020, Marra2019, Scherubl2019, Houzet2019, Repin2019, Sonic2019, Finkelstein2019, WeylDisks, Meyer2017, Eriksson2017} A separate set of proposals aims to realization of Weyl points in devices combining Josephson effect and Coulomb blockade \cite{Fatemi2020, Riwar2020}.

It is important that weak spin-orbit interaction splits the energies of single-quasiparticle states.\cite{Weyl, Yokoyama} Owing to this, the ground state configuration is always a component of a spin doublet in a small finite region around the point and is spin-singlet otherwise.\cite{Yokoyama, Repin2019} The topological singularity still remains since the energies of two singlet states still cross in a point owing to topological protection.

\begin{figure}
\label{fig:setupandspectrum}
\centerline{\includegraphics[width=0.5\textwidth]{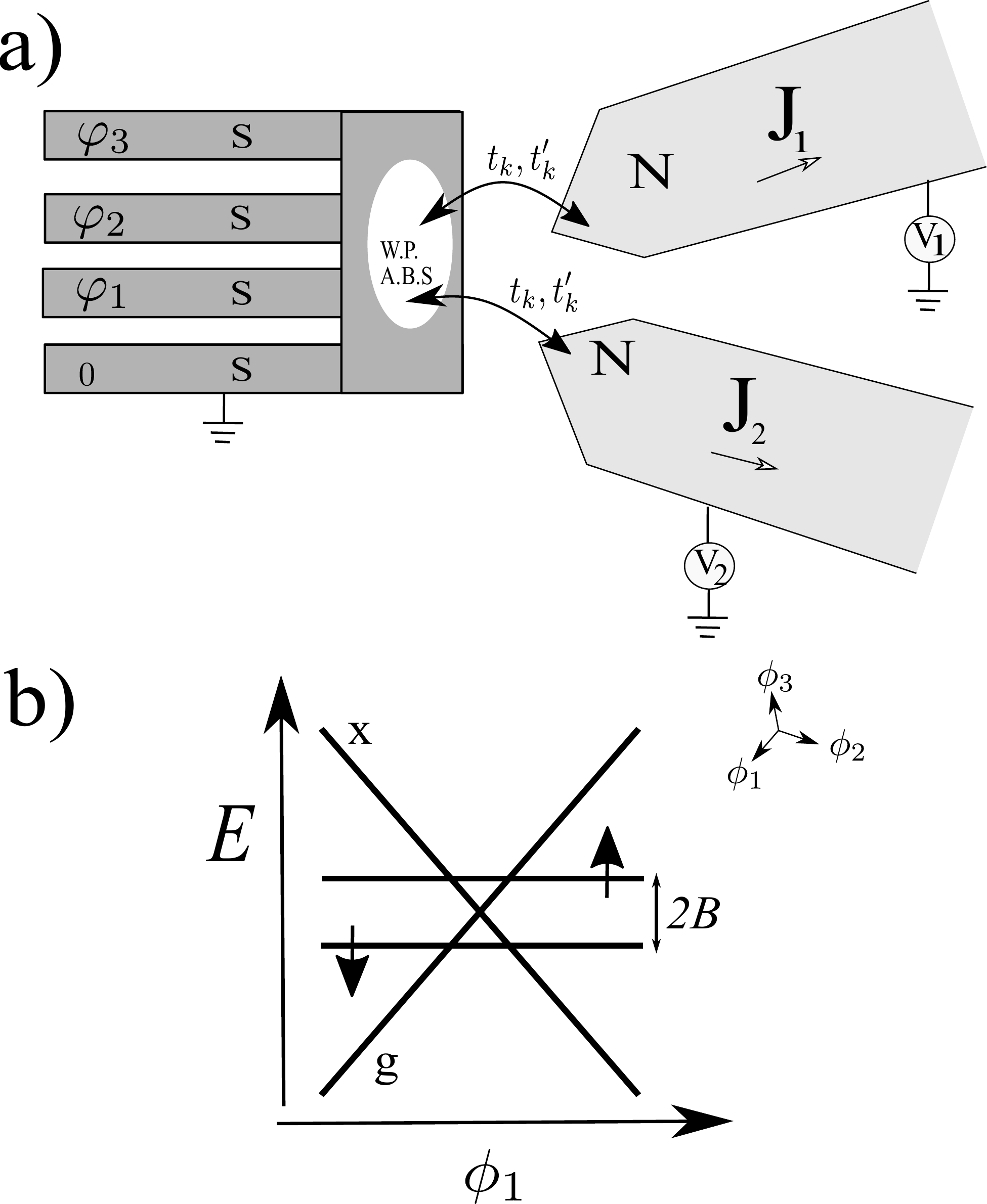}
	}
\caption{a. In a four-terminal superconducting heterostructure, the Andreev states may cross Fermi level in a point - a Weyl point - in 3D parameter space of superconducting phase. The resulting spectrum in the vicinity of the point is isotropic and conical for two singlet states ($x$ and $g$ in the Figure) and flat for doublet states. The doublet states are split by spin-orbit interaction, and one doublet state is ground one in the mere vicinity of the point. b. The setup under consideration. The Andreev bound states near Weyl point (A.B.S.W.P) are tunnel-coupled with the continuous spectrum of the electron states in several normal-metal leads (two are shown in the Figure). The tunnel coupling results in an energy scale $\Gamma$ at which the spectral singularities are smoothed. }
\end{figure}%

In \cite{Repin2019} we have noticed that continuous spectrum above the gap may modify the signatures of topology leading to a non-quantized contribution to the transconductance. The continuous spectrum at low energies shall bring more drastic modification. The most experimentally relevant way to bring a continuous spectrum into play is to couple a system of discrete Andreev levels in the superconducting nanostructure to normal leads. As we will see in detail in this Article, this brings new energy scale $\Gamma$, that is the rate of tunnelling to the leads from a discrete state. Since we are at a point of energy crossing, this small energy scale also implies a small scale in the parameter space: the scale at which the energy splitting matches $\Gamma$.

 We have studied tunnel coupling to discrete normal states in \cite{ChenSpinWeyl} where we propose a Spin-Weyl quantum unit. Importantly, we have found there that the tunnel coupling may break isotropy in the vicinity of the Weyl point. In the context of spintronics, we have recently studied the charge and spin transport in normal leads tunnel-coupled to a Weyl-point superconducting nanostructure. This is essentially the same setup as we consider here. However, in \cite{ChenSpinWeyl} we access the transport in the framework of master equation, that is, assuming that the energy differences of Andreev states exceed much the tunnel energy scale. In this approximation the quantities characterizing the setup retain singularities: the superconducting current has a jump at the point, the normal currents jump at voltages corresponging to the energy levels, the Berry curvature diverges upon approaching the point indicating the point-like topological charge.

In this Article, we investigate the setup at the energy scale $\Gamma$ 
revealing how the above-mentioned singularities are smoothed at this scale. We formulate a generic model of tunneling suitable for many leads that includes isotropy violation.  Technically, the problem at hand is a case of non-equlibrium Green function technique \cite{Keldysh,Rammer} for non-interacting Fermions. However, we chose to present an explicit derivation in terms of Heisenberg equation of motion for the operators of the superconducting current and those of the currents in the normal leads. We compute these quantities for equilibrium, stationary and adiabatic cases. Owing to simplicity of the generic setup under consideration, all results are analytical.

As expected, all singularities are smoothed. We find the maximum derivative of the supercurrent with respect to the controlling phases, that is set by $\Gamma$, and the maximum 
differential conductance in the tunneling currents. An experimentally relevant point is the sharp dependence of tunneling currents in the vicinity of the point in the limit of high voltages and temperatures. This can be used for detection of Weyl points at temperatures that exceed the level splitting.

 We redefine Berry curvature in terms of the response function in the limit of small frequencies. The divergence of the redefined curvature gives the density of topological charge, so we explicitly compute how the point-like topological singularity is spread over the parameter space.

In addition, we evaluate the tunneling currents generated by an adiabatic change of the controlling phases. This is the case of parametric charge pumping\cite{Brouwer, Blaauboer, QuantumTransport}: the result of a change of the controlling phases along a closed contour is a charge transferred to the leads that depends on the contour only. We show that this is a convenient tool for exploration of the vicinity of the Weyl point, including the smoothing of the singularities.  

The structure of the Article is as follows. We formulate the model in Section \ref{sec:model} and perform necessary derivations in Section \ref{sec:derivation}. We evaluate the superconducting currents in equilibrium in Section \ref{sec:equilibrium}. There are no tunneling currents in equilibrium. They arise if the voltages are applied to the leads, and we evaluate these currents for stationary voltages in Section \ref{sec:stationary}. Next, we turn to the adiabatic case computing the response functions in the limit of low frequency.
We redefine Berry curvature, evaluate the response function and the density of topological charge in Section \ref{sec:berry}. The Section \ref{sec:pumping} concentrates on charge pumping to the normal leads. We conclude in Section \ref{sec:conclusions}.

\section{The Model}
\label{sec:model}
We start with the effective Hamiltonian in the vicinity of a Weyl point following \cite{Weyl, Yokoyama, Repin2019}. 

Three independent superconducting phase differences can be regarded as a 3D vector $\vec{\varphi}$. Suppose the Weyl points are situated at $\pm \vec{\varphi}_0$.  In the vicinity of the point at $\vec{\varphi}_0$ we expand $\vec{\varphi} = \vec{\varphi}_0 + \delta\vec{\varphi}$, $|\delta\vec{\varphi}|\ll 1$ and can describe the lowest Andreev bound states by a $2\times 2$ matrix BdG Hamiltonian
\begin{equation}
\hat{H}_{{\rm W}} = \phi_a \hat{\tau}_a; \; \phi_a = M_{ab} \delta \varphi_b,
\end{equation} 
where $\hat{\tau}_a$ is a vector of Pauli matrices. This form suggests convenient coordinates $\vec{\phi}$ for the vicinity of a Weyl point that are linearly related and thus equivalent to $\delta \vec{\varphi}$. We will make use of these coordinates through the paper. In these coordinates of dimension energy, the spectrum is isotropic and conical, $E = \pm |\vec{\phi}|$. The coordinates are thus defined upon an orthogonal transform.


Weak spin-orbit interaction within the nanostructure splits the Andreev states in spin\cite{Yokoyama}, resulting in the following Hamiltonian,
\begin{equation}
\hat{H}^{{\rm W}} = \phi_a \hat{\tau}_a + B_a \hat{\sigma}_a, 
\end{equation}
$\hat{\sigma}_a$ being a vector of Pauli matrices in spin space, and $B_a$ looks like an external magnetic field causing Zeeman splitting. However, $\vec{B} \ne 0$ even in the absence of external magnetic field and represents the effect of the superconducting phase differences on spin orientation. Owing to global time reversibility, the vectors $\vec{B}$ are opposite for opposite Weyl points, $\vec{B}(-\varphi_0)= - \vec{B}(\varphi_0)$. The magnitude of $\vec{B}$ can be estimated as the superconducting energy gap $\Delta$ times a dimensionless factor characterizing the weakness of the spin-orbit interaction. For a concrete number in mind, we can take $B \simeq 0.1 \Delta \simeq 0.2 meV$ which corresponds to niobium. If there is an external magnetic field, it adds to $\vec{B}$. We note however that our estimation of $B$ is about $3 T$, so it requires a significant field to change it.

To represent  the Hamiltonian in the second-quantization form, we introduce quasiparticle annihilation operators $\hat{\gamma}_\sigma$ and associated 4-component Nambu bispinors 
$\gamma_\alpha$, where $\alpha = (i,\sigma)$ combines spin and Nabmu index $i=e,h$, $\bar{\gamma}_{i,\sigma} \equiv (\hat{\gamma}_\sigma, -\sigma \hat{\gamma}_{-\sigma})$ to recast it to the standard form,
\begin{equation}
H_{{\rm W}} = \frac{1}{2} \bar{\gamma}^\dagger_\alpha H^{{\rm W}}_{\alpha \beta}  \bar{\gamma}_\beta.
\end{equation}
We note that $\gamma^\dagger_{i,\sigma} = -\sigma \gamma^{-i,-\sigma}$
This gives an isotropic spectrum which depends only on $\phi \equiv |\vec{\phi}|$.(see Fig. \ref{fig:setupandspectrum} a) The energies are $E = \pm \phi$ for two spin-singlet states, ground one $|g\rangle$, and excited one $|x\rangle$, and $E = \pm B$ for two components of the spin doublet $|\uparrow\rangle$, $|\downarrow\rangle $. The energies of the split doublet exhibit no singularity nor phase dependence in the vicinity of the Weyl point, while the spin-singlet states retain the conical spectrum. 

The ground state is magnetic ($|\downarrow\rangle$) in a narrow vicinity of the Weyl point, namely, at $|\phi|<B$ and spin-singlet otherwise. (Fig. \ref{fig:setupandspectrum} a) 


We will need the current operators in 3 superconducting leads. They are given by the derivatives of the Hamiltonian with respect to the phases,\cite{QuantumTransport}
\begin{equation}
I_a = \frac{2e}{\hbar}\frac{\partial \hat{H}_W}{\partial \varphi_a} = \frac{2e}{\hbar} M_{ab} \tilde{I}_b;
\end{equation}
\begin{equation}
\label{eq:currents}
\tilde{I}_a \equiv \frac{1}{2} \gamma^\dagger_\alpha \tau^a_{\alpha \beta} \gamma_\beta
\end{equation}
Since there is a trivial linear relation between $I_a$ and ${\tilde I}_a$, we will futher concentrate on the dimensionless quantities ${\tilde I}_a$.

Let us bring in the coupling with the continuous spectrum of electron states in several leads (Fig. \ref{fig:setupandspectrum}).
We will describe the leads with a usual free-fermion Hamiltonian
\begin{equation}
\hat{H}_{\rm leads} = \sum_{k} E_{k}   \hat{d}^{\dag}_{k,\sigma} \hat{d}^{a}_{k,\sigma}
\end{equation}
where  $k$ labels the  states of the quasi-continuous spectrum in the leads, $d_k$ are the corresponding electron annihilation operators, $E_k$ are the corresponding energies. The states $k$ are distributed over the leads, those are  labelled with $a$. We characterize a general non-equilibrium state of the leads with the energy-dependent filling factors $f_a(E)$ such that 
\begin{equation}
\label{eq:fillings}
\langle \hat{d}^{\dag}_{k,\sigma} \hat{d}_{k,\sigma} \rangle = f_a(E_{k}) \ for \ k \in a.
\end{equation} 

The crucial part of the Hamiltonian is the tunnelling between the electron states in the leads and the Andreev state in the nanostructure. We will keep it in the most general form,
\begin{equation}
\hat{H}_{{\rm T}} = \sum_{k,\sigma} \left( t_{k} \hat{\gamma}^\dag_\sigma  - t'_{k}\sigma \hat{\gamma}_{-\sigma}\right)\hat{d}_{k,\sigma} + h.c.
\end{equation}
not specifying the spin-independent tunnel amplitudes $t_{k}, t'_{k}$. In the course of the derivation, we will see  which combinations of the amplitudes are the relevant parameters of the model.
 It is convenient to present the Hamiltonian in the form of Nambu spinors
\begin{equation}
2\hat{H}_{{\rm T}} = \sum_{k} \gamma^\dag_{\alpha} T^{\alpha\beta}_{k} d^{\alpha}_{k} + h.c.
\end{equation}
where the matrix $T^{\alpha\beta}$ depends on the Nambu index only,
\begin{equation}
T_{k} = \begin{pmatrix}
t^{k}& t_{k}^{\prime *}\\
t_{k}^{\prime}&-t_{k}^{*}
\end{pmatrix}
\end{equation}
With this, we derive the operators of the current to a normal lead $a$
\begin{align}
J_a = e\sum_{k \in a,\sigma} i \left( t_{k} \hat{\gamma}^\dag_\sigma  - t'_{k}\sigma \hat{\gamma}_{-\sigma}\right)\hat{d}_{k,\sigma} + h.c;\\
J_a = \frac{ie}{2} \sum_{k \in a} \gamma^\dag_{\alpha} (T^{\alpha\beta}_{k}\tau_3)^{\alpha\beta} d^{\alpha}_{k} + h.c. 
\end{align}
\section{Derivation}
\label{sec:derivation}
The derivation of expressions for the currents in superconducting and normal leads can be accomplished by standard methods of superconducting non-equilibrium Keldysh Green functions \cite{Keldysh,Rammer,Baranski}. However, for the sake of comprehensibility we give here an explicit derivation from scratch. This is easy for the system under consideration and makes explicit the transition from quasi-continuous to continuous spectrum in the leads. 

Let us write down the Heisenberg evolution equations for the operators $\hat{\gamma}^\alpha, \hat{d}^\alpha_{k,\sigma}$ governed by the total Hamiltonian $\hat{H} = \hat{H}_{{\rm W}} + \hat{H}_{\rm leads} + \hat{H}_{\rm T}$. We use bold-face notations for bispinors and "check" for the corresponding $4\times4$ matrices. In these notations,
\begin{align}
\label{eq:gamma}
i\dot{\boldsymbol{\gamma}} &= \check{H}_{{\rm W}} \boldsymbol{\gamma}+\sum_{k}\check{T}_{k} \boldsymbol{d}_{k}
\\
i\dot{\boldsymbol{d}}_{k} &= E_{k} \check{\tau}_3 \boldsymbol{d}_{k}+\check{T}^{\dag}_{k}\boldsymbol{\gamma}
\end{align}
Here, we implicitly assume a time-dependence of $H^{{\rm W}}$. 
Solving equations for each of $\hat{d}$ gives
\begin{equation}
\boldsymbol{d}_{k} (t) = e^{-i E_{k} \check{\tau}_3 t} \boldsymbol{d}^{0}_{k} + \int dt'\check{g}_{k} (t,t')\check{T}^\dag_{k}\boldsymbol{\gamma}(t') \
\end{equation} 
where 
\begin{equation}
\check{g}_{k} (t,t') = -i e^{-i E_{k}\check{\tau}_3(t-t')} \Theta(t-t').
\end{equation}
Here, $\boldsymbol{d}^0$ describes the state of the leads. 
We substitude this to Eq. \ref{eq:gamma} to obtain a closed equation for $\boldsymbol{\gamma}$ and express it in terms of $\boldsymbol{d}^0$:
\begin{equation}
\label{eq:gammaG}
\boldsymbol{\gamma}(t) = \int dt' \check{G}(t,t') \sum_{k}\check{T}_{k} e^{-i E_{k} \check{\tau}_3 t'}\boldsymbol{d}^{0}_{k}
\end{equation}
where we have introduced the advanced Green function defined as
\begin{equation}
[i\partial_t-\check{H}_{{\rm W}}]\check{G}(t,t')-\int dt'' \check{\Sigma}(t-t'')\check{G}(t'',t')  = \delta(t-t')
\end{equation}
where the self-energy  $\check{\Sigma}$ reads
\begin{equation}
\check{\Sigma}(t,t') = \sum_{k} \check{T}_k \check{g}_{k}(t,t') \check{T}^\dag_{k}
\end{equation}

We substitute the expression (\ref{eq:gammaG}) to the expressions for the current operators (\ref{eq:currents}) and average over the non-equilibrium state of the leads using Eq. \ref{eq:fillings}.
This yields
\begin{equation}
\label{eq:Ia}
\langle\tilde{I}_a\rangle = \frac{1}{2} \int dt' dt'' {\rm Tr}[\check{\tau}_a \check{G}(t,t') \check{F}(t',t'') \check{\bar{G}}(t'',t)]
\end{equation}
where $\check{\bar{G}}(t,t') \equiv \check{G}^\dag (t',t)$ and 
\begin{equation}
\check{F} = \check{T}_{k} \begin{pmatrix}
f_{k}e^{iE_{k}(t'-t)}& 0\\
0& (\bar{f}_{k}) e^{iE_{k}(t-t')} 
\end{pmatrix}\check{T}^\dag_{k} 
\end{equation} 
Here and further on, $\bar{f}_k \equiv 1-f_k$.
In a similar way, we derive the averages of the currents in the normal 
leads. They read:
\begin{align}
\label{eq:Ja}
\langle J_a(t)\rangle = e\int dt_1 dt_2 dt_3 {\rm Tr}[\check{M}_a(t,t_1) \check{G}(t_1,t_2) \nonumber \\
\check{F}(t_2,t_3)\check{\bar{G}}(t_3,t)] +
\int dt_1 \left( {\rm Tr}[\check{D}_a(t,t') \check{G}(t,t')] + h.c. \right).
\end{align}
Here, we define 
\begin{equation}
\check{M}_a = -\frac{1}{2}\sum_{k \in a} \check{T}_{k} \tau_3 e^{-i E_k \tau_3(t-t')} \check{T}_k^\dag;
\end{equation}
\begin{equation}
\check{D}_a (t,t') = \frac{-i}{2}\sum_{k \in a} \check{T}_k\tau_3 \begin{pmatrix}
f_{k}e^{iE_{k}(t'-t)}& 0\\
0& f_{k} e^{iE_{k}(t-t')} 
\end{pmatrix}\check{T}^\dag_k.
\end{equation}

So far, the expressions are valid for any spectrum in the normal lead, either quasi-continuous or continuous. Let us now specify to continuous spectrum. For this, we define the following combinations of tunnel amplitudes in each lead:
\begin{equation}
\Gamma_a (E) = \sum_{k \in a} (|t_{k}|^2+|t'_{k}|^2) \delta(E - E_{k});
\end{equation} 
\begin{equation}
\vec{\Gamma}_a(E) = \sum_{k \in a} (2{\rm Re}(t'_{k} t^*_{k}),2{\rm Im}(t'_{k} t^*_{k}), |t_{k}|^2-|t'_{k}|^2) \delta(E - E_{k})
\end{equation}
All the constituents of the expressions for the operators can be expressed through $\Gamma_a(E), \vec{\Gamma}_a(E)$. Those are thus the actual parameters of our model. The continuous spectrum is implemented by assumption that $\Gamma_a(E), \vec{\Gamma}_a(E)$ are continuous and smooth functions of energy. Moreover, a convenient and relevant assumption is that these functions vary at an energy scale that exceeds by far that of the Weyl point. In this case, the energy dependence can be disregarded and $\Gamma_a, \vec{\Gamma}_a$ are taken at zero energy.

Let us see how $\check{\Sigma}$, $\check{F}$, $\check{M}_a$ and $\check{D}_a$ are simplified under these assumptions. In energy representation, the self-energy becomes
\begin{equation}
\check{\Sigma}(\epsilon) =\frac{1}{4\pi}\sum_{\pm} \left(\Gamma(E) \pm \vec{\Gamma}(E) \cdot \vec{\check{\tau}}\right) \frac{1}{\epsilon \mp E -i0}
\end{equation}
where $\Gamma, \vec{\Gamma} \equiv \sum_a \Gamma_a, \vec{\Gamma}_a$. The Hermitian part of $\check{\Sigma}$ in the limit $\epsilon$ adds a constant term to $H_{{\rm}}$ and therefore describes a shift, or renormalization of the Weyl point position in the space of three pahses due to tunneling,
\begin{equation}
\delta \phi = -\int dE\frac{\vec{\Gamma}(E)}{E}.
\end{equation} 
We will disregard this irrelevant redifinition of the Weyl point position.
The anti-Hermitian part of the self-energy is more imortant describing the decay of discrete states into the continuous spectrum,
\begin{equation}
\check{\Sigma} = \frac{1}{4}\sum_{\pm} \left(\Gamma(\pm \epsilon) \pm  \vec{\Gamma}(\pm \epsilon) \cdot \vec{\check{\tau}}\right) \approx \frac{\Gamma}{2}
\end{equation}
where the limit of small $\epsilon$ has been implemented in the last equality.
The matrices $\check{F}, \check{D}_a$ bring the information about the filling factors in the leads and are expressed as
\begin{eqnarray}
\check{F} = \sum_a \Gamma_a f^{+}_a + \vec{\Gamma}_a \cdot \vec{\check{\tau}} f^{-}_a \\
\check{D}_a = -\frac{i}{2} \left[ \vec{\Gamma}_a \cdot \vec{\check{\tau}} f^{+}_a + \Gamma_a f^{-}_a \right].\\
f^{\pm}(\epsilon) \equiv \frac{f_a(\epsilon) \pm \bar{f}_a(-\epsilon)}{2}
\end{eqnarray}
Finally, $\check{M}_a= -\vec{\Gamma}_a \cdot \vec{\check{\tau}}/2$.
With this, the terms with $\check{M}_a$ in Eq. \ref{eq:Ja} are related to superconducting currents,
\begin{equation}
\label{eq:Ja-simple}
\langle J_a\rangle = -\vec{\Gamma}_a \cdot \vec{\tilde{I}}+
\int dt_1 \left( {\rm Tr}[\check{D}_a(t,t') \check{G}(t,t')] + h.c. \right)
\end{equation}
From now on, we will denote the expectation values of the currents simply as $J_a$, $\vec{\tilde{I}}$. 

\section{Currents in Equilibrium}
\label{sec:equilibrium}
In equilibrium and stationary state, the Green functions are diagonal in energy representation,
\begin{equation}
\label{eq:Green}
\check{G}, \check{\bar{G}} = \frac{1}{\epsilon - \check{H}_{{\rm W}} \mp i\frac{\Gamma}{2}}.
\end{equation}
There is also a convenient relation
\begin{equation}
\label{eq:convenient}
i(\check{G}^{-1}-\check{\bar{G}}^{-1}) = \Gamma 
\end{equation}
We note that in equilibrium $f(\epsilon)=\bar{f}(-\epsilon)$ and filling factors in all leads correspond to Fermi distribution at zero chemical potential, $f_a(\epsilon) = f_F(\epsilon)$.
With this, $\check{F} = \Gamma f_F$. Invoking Eq. \ref{eq:convenient}, we prove
\begin{equation}
\label{eq:relationeq}
\check{G}\check{F}\check{\bar{G}} = -i f_F(\check{G} - \check{\bar{G}})
\end{equation}
and the currents are expressed as
\begin{equation}
\vec{\tilde{I}} = -i\int \frac{d\epsilon}{2\pi} {\rm Tr}[\vec{\check{\tau}} (\check{G} -\check{\bar{G}}) f_F(\epsilon)]
\end{equation}

Let us first recognize that the equilibrium super currents are expressed from the derivatives of free energy with respect to $\vec{\phi}$. For an isolated superconducting nanostructure, that is, in the limit $\Gamma \ll B, \phi$, and at zero temperature, the ground state energy is given through the positive energies of Andreev bound states,
\begin{equation}
E_g = - \frac{1}{2} \sum_i E_i \Theta(E_i)
\end{equation}
For the nanostructure under consideration, the Andreev bound states are $E_{\sigma, \pm} = B\sigma \pm \phi$ and the currents in this limit read
\begin{equation}
\vec{\tilde{I}} = - \vec{n} \Theta(\phi - B)
\end{equation}
The current has a cusp: that is, its derivative with respect to $\phi$ diverges in a point. This divergence may be in principle used for finding the Weyl point and is smoothed at the scale of $\Gamma$.

\begin{figure}
\label{fig:current}
\centerline{\includegraphics[width=0.5\textwidth]{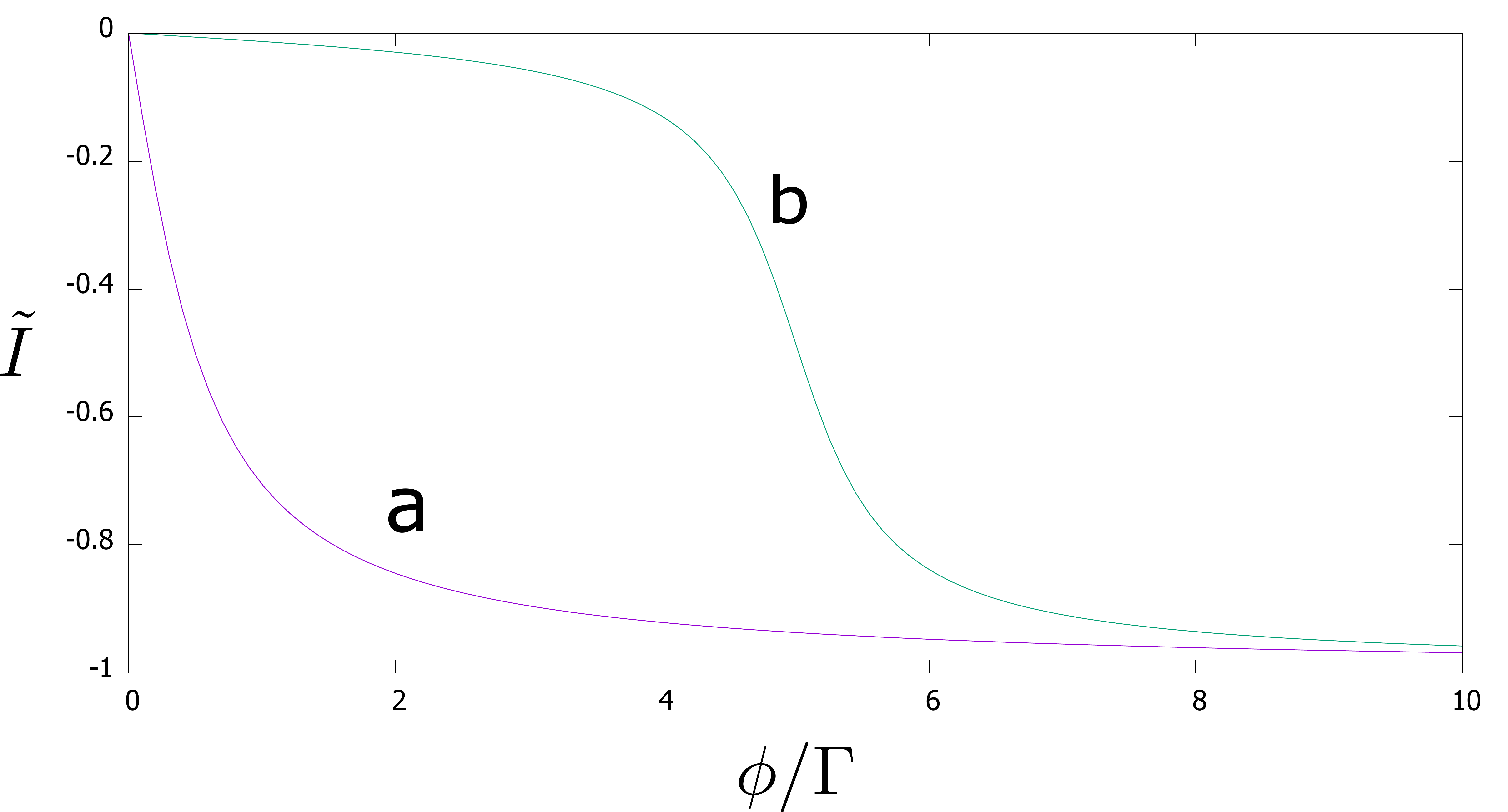}
	}
\caption{Smoothing of the superconducting current singularity at the scale of $\Gamma$. The curve a. corresponds to $B\ll \Gamma$, while the curve b. to $B=5\Gamma$. }
\end{figure}%

At finite $\Gamma$, the Andreev energies correspond to the poles of the Green functions. Their poles are shifted by $\pm i \Gamma/2$ from the real axis. The currents are expressed through the phases of the pole positions $\xi_{\sigma, \pm} \equiv \arctan(2(B\sigma \pm \phi)/\Gamma)$,
\begin{equation}
\vec{\tilde{I}} = \frac{\vec{n}}{2\pi} \sum_\sigma (\xi_{\sigma,-} - \xi_{\sigma,+})
\end{equation} 
The cusps are smoothed by a finite $\Gamma$ (see Fig. \ref{fig:current}). The maximum derivative with respect to $\phi$ is now finite and is of the order of $\Gamma^{-1}$:
\begin{equation}
\frac{\partial {\tilde I}}{\partial \phi} = \frac{2}{\pi \Gamma} \ {\rm for} \ B \ll \Gamma, \ \ \frac{1}{\pi \Gamma} \ {\rm for} \ B \gg \Gamma.
\end{equation}

In equlibrium, we expect no currents to normal leads.
Indeed, if there were currents, one could extract energy from the equilibrium system by applying voltages to the normal leads. Technically, two terms in Eq. \ref{eq:Ja-simple} cancel each other upon applying the relation (\ref{eq:relationeq}).

\section{Stationary currents}
\label{sec:stationary}
Now we turn to the case of non-equilibrium filling factors in the leads still assuming stationary Weyl point Hamiltonian. The currents are given by Eqs. \ref{eq:Ja-simple}, \ref{eq:Ia} with energy-diagonal Green functions (\ref{eq:Green}).  To keep the formulas simple, we will specify to differential conductances at vanishing temperature. The voltages in the leads only change the filling factors, at vanishing temperature $\partial f_a /\partial_{eV_a} = \delta(\epsilon - eV_a)$, that is, the differential conductances are contributed by the specific energies $\epsilon = \pm eV_a$ only. 

\begin{figure}
\label{fig:tildeIder}
\centerline{\includegraphics[width=0.5\textwidth]{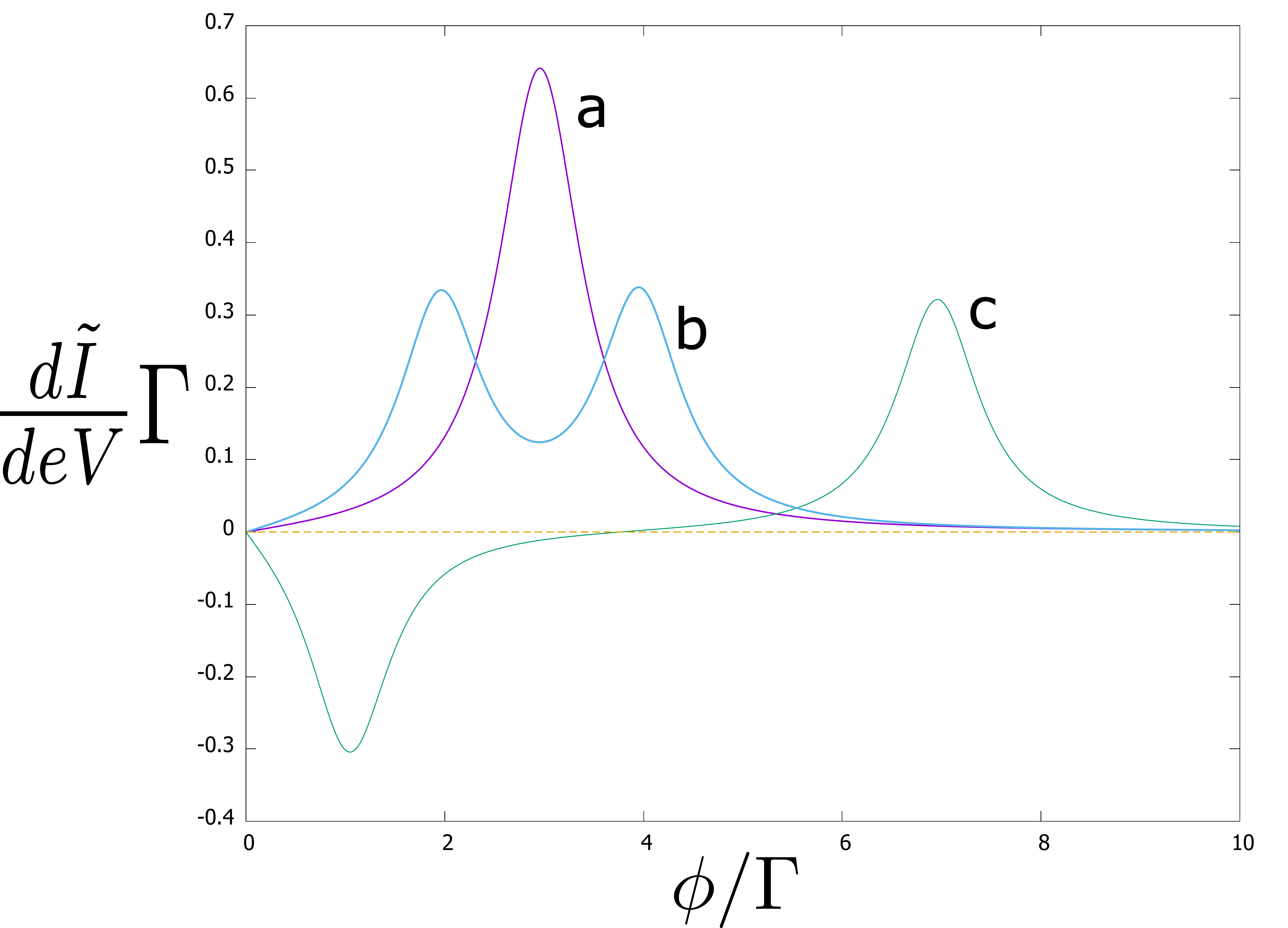}
	}
\caption{The voltage derivative of the superconducting current. There is a single lead, $\vec{\Gamma}=0$, we set $\phi =3.0$. The spin splitting $B$ is set to 0, $\Gamma$, $4\Gamma$, for the curves a,b,c respectively.}
\end{figure}

For the derivatives of supercurrents, we have
\begin{align}
\label{eq:tildeIder}
2\pi \frac{\partial\vec{I}}{\partial{eV_a}} = \vec{\phi} \Gamma_a K_{o}(eV_a) + \nonumber\\ 
(2 \vec{\phi} \cdot \vec{\Gamma}_a) \vec{\phi} +(\vec{\phi} \times \vec{\Gamma}_a) )K_{e}(eV_a) +\vec{\Gamma}_a K_3(eV_a)
\end{align}
where the functions $K_{o,e,3}$ are defined as ($K^{-1}_\sigma \equiv ((\epsilon-B\sigma)^2 - \Gamma^2/4 -\phi^2)^2 +\Gamma^2(\epsilon-B\sigma)^2$):
\begin{align}
\label{eq:Kdefs}
K_o &=& 2\sum_\sigma (\epsilon-B\sigma) K_\sigma;  K_e = \sum_\sigma K_\sigma; \\
K_3 &=&\sum_\sigma ((\epsilon-B\sigma)^2 + \Gamma^2/4 -\phi^2) K_\sigma
\end{align}
We note that 
\begin{align}
\label{eq:Kint}
\int_0^\infty d \epsilon K_o &=& \frac{2(\arctan(\phi+B)+\arctan(\phi-B))}{\Gamma \phi};\\ \int_0^\infty d \epsilon K_e &=& \frac{\pi}{\Gamma(\Gamma^2/4+\phi^2)}; \\
\int_0^\infty d \epsilon K_3 &=&\frac{\pi \Gamma}{2(\Gamma^2/4+\phi^2)} 
\end{align}
The derivatives are illustrated in Fig. \ref{fig:tildeIder} for a single lead and simple case $\vec{\Gamma}=0$. They peak at the positions of resonant levels $eV =\phi+B$, $|\phi -B|$. The peak width is of the order of $\Gamma$. For singlet ground state (the curves $a$,$b$ the finite current at zero voltage falls to zero in one or two steps. 
For the doublet ground state, the current that is small at zero voltage rises at the first and drops at the second resonant level. 

The differential conductances in the normal leads are given by:
\begin{align}
\frac{\partial J_a}{\partial e^2 V_b} = - \vec{\Gamma}_a \cdot \frac{\partial \vec{\tilde{I}}}{\partial eV_b}+ \nonumber\\
\frac{\Gamma\delta_{ab}}{2\pi} \left( \Gamma_a (K_3(eV_a) + 2\phi^2 K_{e}(eV_a)) + (\vec{\Gamma}_a \cdot \vec{\phi})K_{o}\right)
\end{align}

\begin{figure}
\label{fig:zerovoltage}
\centerline{\includegraphics[width=0.5\textwidth]{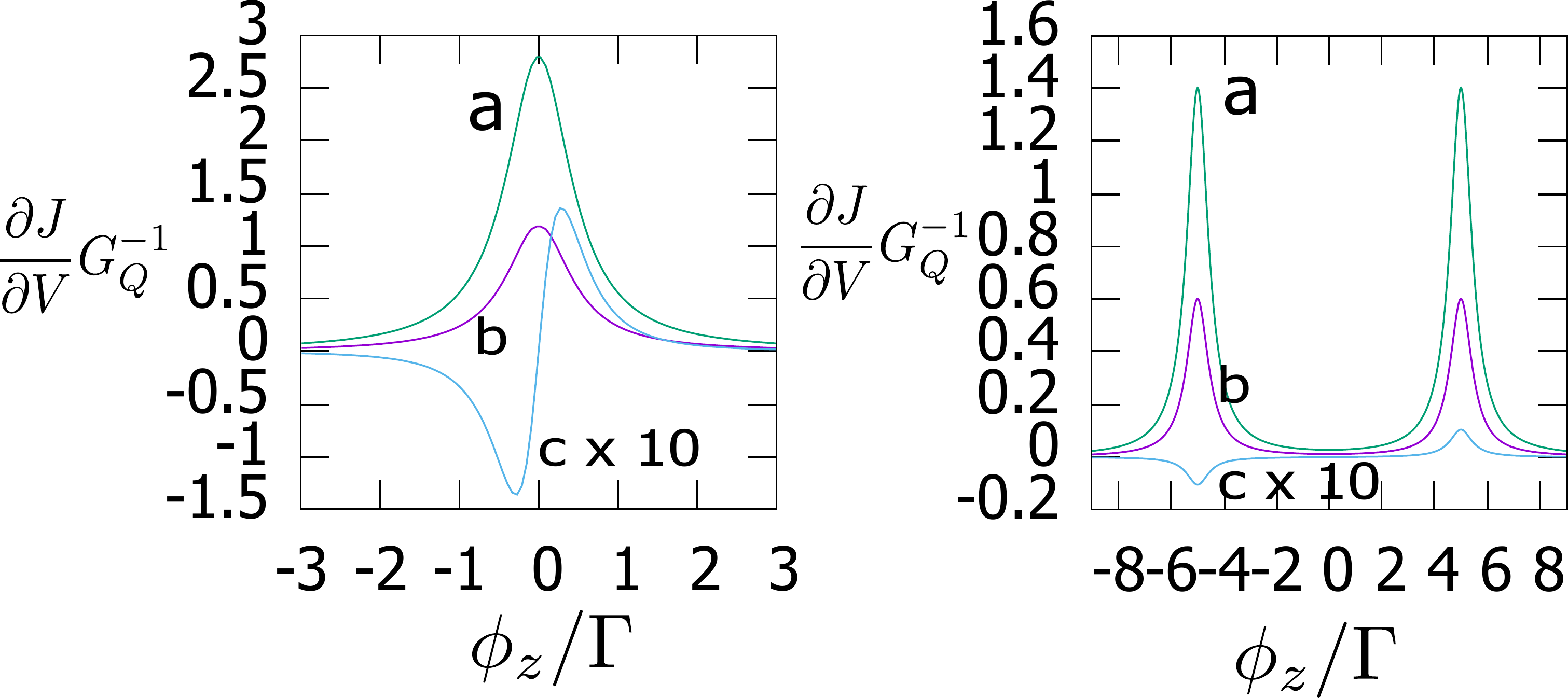}
	}
\caption{An example of zero-voltage conductances. There are two leads, $\Gamma_1=0.7\Gamma$, $\Gamma_2=0.3 \Gamma$, $\vec{\Gamma}_1 \parallel x$,$\vec{\Gamma}_2 \parallel y$, the plots are for $\vec \phi$ in z-direction. The curves a,b,c, correspond two $G_{11},G_{22},G_{12}$. The transconductance is antisymmetric in this case, $G_{12}=-G_{21}$. Left pane: $B\ll \Gamma$, right pane: $B=5 \Gamma$. The vertical scale of the curve c is increased by a factor of 10. }
\label{system}
\end{figure}%
We plot in Fig. \ref{fig:zerovoltage} an example of zero-voltage conductances $G_{11},G_{22},G_{12}$ for two leads. The diagonal conductances peak when the resonant levels are at zero energy, $|\phi -B|=0$. The peak widths are of the order of the conductance quantum $G_Q \equiv e^2/\hbar pi$. An interesting feature is a Hall-like antisymmetric transconductance $G_{12} = -G_{21}$. It incorporates the effects of vector parts of $\Gamma$ in two leads, $G_{12} \propto \vec{\phi}\cdot (\vec{\Gamma}_1 \times \vec{\Gamma}_2 )$ and changes sign if $\vec{\phi} \to - \vec{\phi}$. 

\begin{figure}
\label{fig:dc}
\centerline{\includegraphics[width=0.4\textwidth]{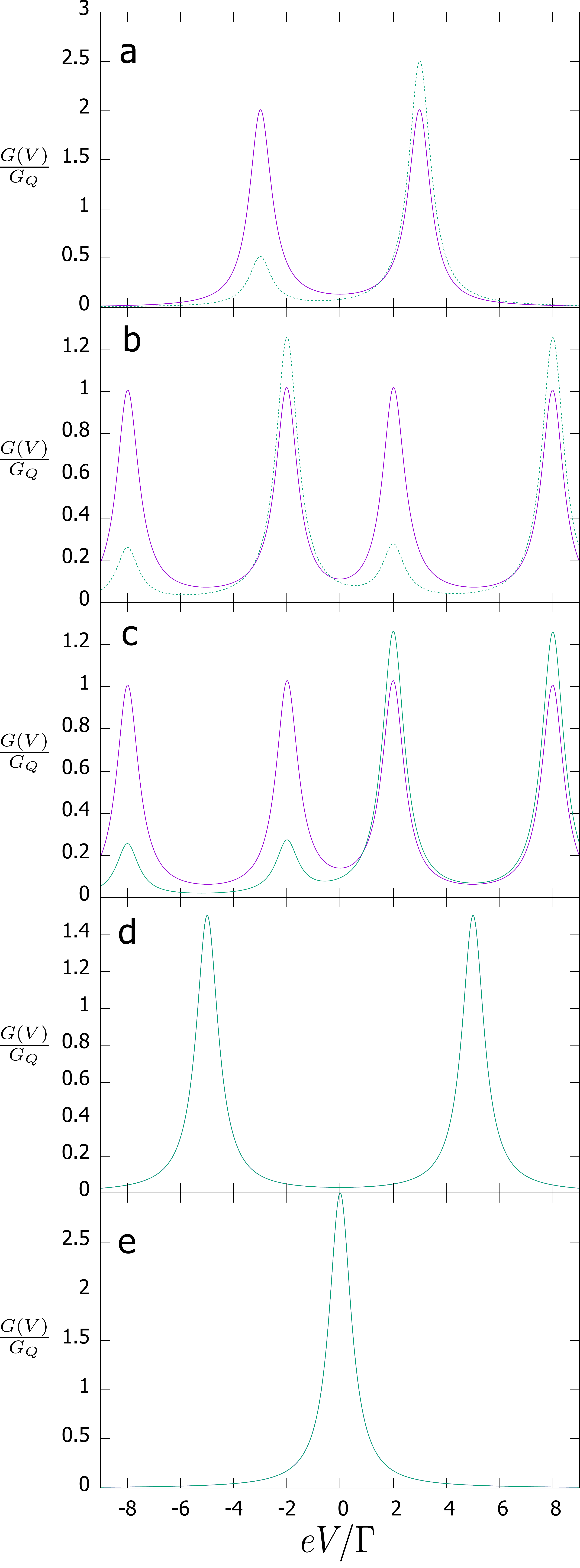}}
\caption{Differential conductance for the case of a single lead.
$|\vec{\Gamma}| = \Gamma/2$ was taken. The solid curves correspond to $\vec{\phi} \perp \vec{\Gamma}$, the conductance is even in $V$ and $\phi$.  The dashed curves correspond to $\vec{\phi} \parallel \vec{\Gamma}$, and $G(V,\phi)=G(-V,-\phi)$. The parameters are: a. $B=0$, $\phi=3 \Gamma$; b. $B=5 \Gamma$, $\phi=3 \Gamma$; c. $B=3 \Gamma$, $\phi=5 \Gamma$; d. $B=5 \Gamma$, $\phi=0$; e. $B=0$, $\phi=0$.  The perpendicular and parallel conductances coincide for the last two plots, since $\phi=0$.}
\end{figure}

For finite-voltage conductance, we restrict ourselves to the case of a single lead. The example for $|\vec{\Gamma}| = \Gamma/2$ is given in Fig. \ref{fig:dc}. The peaks of differential conductance are situated at $eV = |\phi \pm B|$, their width being of the order of $\Gamma$. The peak values are of the order of $G_Q$. The vector part of $\Gamma$ brings anisotropy and asymmetry of conductances with respect to voltage and $\vec{\phi}$. 

At high voltages $eV \gg \Gamma, \phi, B$ applied, the current in the normal lead saturates at finite value $J_\infty$, as it is expected for the transport via resonant levels. We note a peculiar feature: this current retains the dependence on $\phi$ and its direction, this dependence is smoothed at the small scale of $\phi \simeq \Gamma$ only. Using the relations (\ref{eq:tildeIder}),(\ref{eq:Kdefs}),(\ref{eq:Kint}), we obtain 
\begin{equation}
\label{eq:Jinf}
J_\infty/e = \Gamma - \frac{(\vec{\Gamma} \cdot \vec{\phi})^2 +(\vec{\Gamma})^2 \Gamma^2/4}{\Gamma(\phi^2 +\Gamma^2/4)} 
\end{equation} 
This feature survives rather high temperatures $\phi \ll k_B T \ll eV$ at which the thermal equilibration eventually cancels the superconducting currents near the Weyl point. This makes the feature highly proficient for experimental detection of Weyl points in a practical situation where the finite temperature prevents the detection through the supercurrent. One would look at the variation of the tunnel current under variation of $\phi$ to find a signal that is concentrated near the point and shows anisotropy defined by Eq. \ref{eq:Jinf}. The maximum derivative
for $\vec{\phi} \perp \Gamma$  
\begin{equation}
\frac{\partial J}{\partial \phi} = \frac{e}{\hbar} \frac{\vec{\Gamma}^2}{\Gamma^2}  
\end{equation}  
does not depend on the strength of the tunnel coupling, this guarantees a big amplitude of the detection signal.

\section{Redifinition of Berry curvature and density of topological charge}
\label{sec:berry}
In this Section, we consider adiabatic case. We assume equilibrium filling factor in the leads and concentrate on the case of vanishing temperature. If we change the control phases slowly,the superconducting currents acquire a correction proportional to time derivatives of the phases:
\begin{equation}
\tilde{I}^\alpha (t) = \tilde{I}^\alpha(\vec{\phi}(t)) + 
B^{\alpha \beta}(\vec{\phi}) \dot{\phi}^\beta
\end{equation}
Thereby we define a tensor response function $B^{\alpha\beta}$. The symmetric part of this tensor defines the dissipation in the course of the slow change of the phases,
\begin{equation}
\frac{dE}{dt} = \dot{\phi}^\alpha B^{\alpha \beta}(\vec{\phi})\dot{\phi}^\beta.
\end{equation}
If the system under consideration is gapped, 
the dissipative part is absent, while the antisymmetric part of the response function gives the Berry curvature of the ground state of the system (see e.g. \cite{Weyl}) 
\begin{equation}
B^{\alpha \beta}= 2 {\rm Im} \langle \partial_\alpha \Psi|\partial_\Psi \Psi \rangle.
\end{equation}
It is convenient to introduce a pseudovector of Berry curvature $B^\alpha = e^{\alpha \beta \gamma}$ For the superconducting Weyl point, the Berry curvature has been evaluated in \cite{Weyl,Repin2019}. For the singlet ground state, and in the coordinates in use it assumes the standard expression $\vec{B} =\vec{\phi}/(2 \phi^2)$. The flux of $\vec{B}$ through a surface enclosing the origin is $2\pi$ manifesting a unit point-like topological charge at the origin. However, $\vec{B}=0$ at $\phi <B$ where the ground state is doublet. The continuity of the ground state is broken at $\phi = B$ and topological consideration that guarantees a divergentless $\vec{B}$ cannot be applied anymore.

We evalute $B^{\alpha\beta}$ for the setup under consideration making use of Eq. \ref{eq:Ia}. Given a modulation of the Hamiltonian $\check{\delta H}$ oscillating at frequency $\omega$, the response of the currents oscillating at the same frequency can be represented as 
\begin{align}
\tilde{I}^\alpha_\omega = \int \frac{d \epsilon}{2\pi} \frac{1}{2} {\rm Tr} [ \check{\tau}^\alpha (\check{G}_{\epsilon+\omega} \check{\delta H} \check{G}_\epsilon \check{F}_\epsilon \check{\bar{G}}_\epsilon +\\
 \check{G}_\epsilon \check{F}_\epsilon \check{\bar{G}}_\epsilon \check{\delta H} \check{\bar{G}}_{\epsilon-\omega})].
\end{align}
We obtain $B^{\alpha\beta}$ by substituting $\check{H} = \delta\phi^\alpha \check{\tau}^\alpha$ and taking the limit $\omega \to 0$. This is valid for $\omega \ll \Gamma$. We assume vanishing temperature when integrating over the energy.

\begin{figure}
\label{fig:diss}
\centerline{\includegraphics[width=0.5\textwidth]{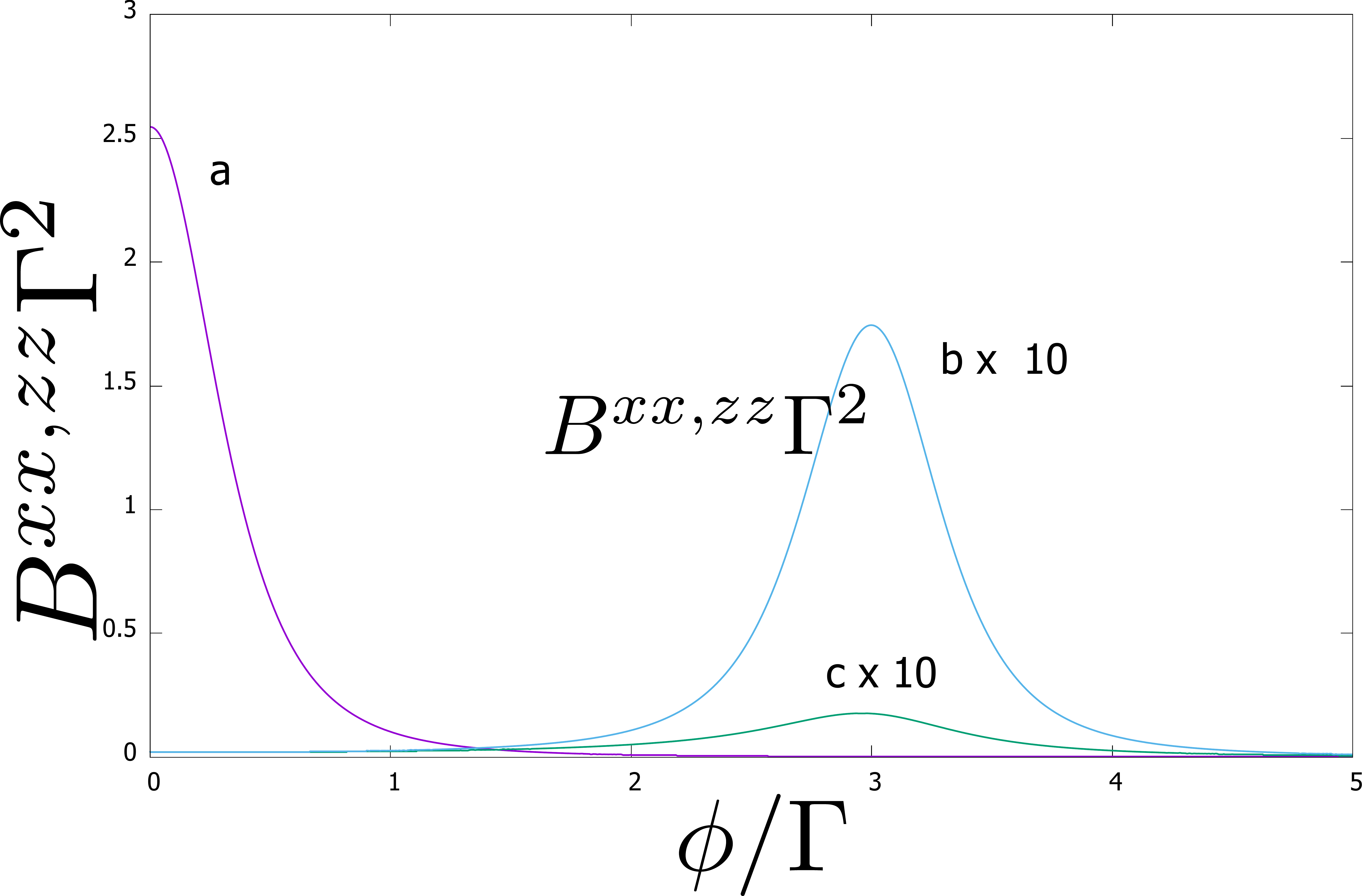}}
\caption{Dissipative part of the response function. We assume $\phi \parallel z$ and plot $B^{zz}$, $B^{xx}=B^{yy}$. Curve a: $B=0$, $B^{zz}=B^{xx}$. Curves b,c: $B=3$
}
\end{figure}

To present the answers in a compact form, we introduce a convenient expression $K \equiv (\phi^2-B^2+\Gamma^2/4)^2 +B^2 \Gamma^2$. The dissipative part of the response function reads:
\begin{equation}
B^{\alpha \beta} =\frac{\Gamma^2}{2\pi K} \left(\delta_{\alpha \beta} +\frac{\phi^\alpha\phi^\beta B^2}{K}\right) 
\end{equation} 
It is plotted in Fig. \ref{fig:diss} for two values of magnetic field. 
We note that the dissipative part at small $\Gamma$ is proportional to $\Gamma^2$ except $\phi=B$ This is because the dissipation requires an excitation of an electron-hole pair in the normal leads, which is a second-order tunneling process \cite{QuantumTransport}. At the resonance threshold $\phi=B$, and $B\gg \Gamma$, the dissipative part of the response function is strongly anisotropic: it is  $\simeq \Gamma^{-2}$ for the direction $\parallel \vec{\phi}$ and $\simeq B^{-2}$ otherwise. 


\begin{figure}
\label{fig:berry}
\centerline{\includegraphics[width=0.5\textwidth]{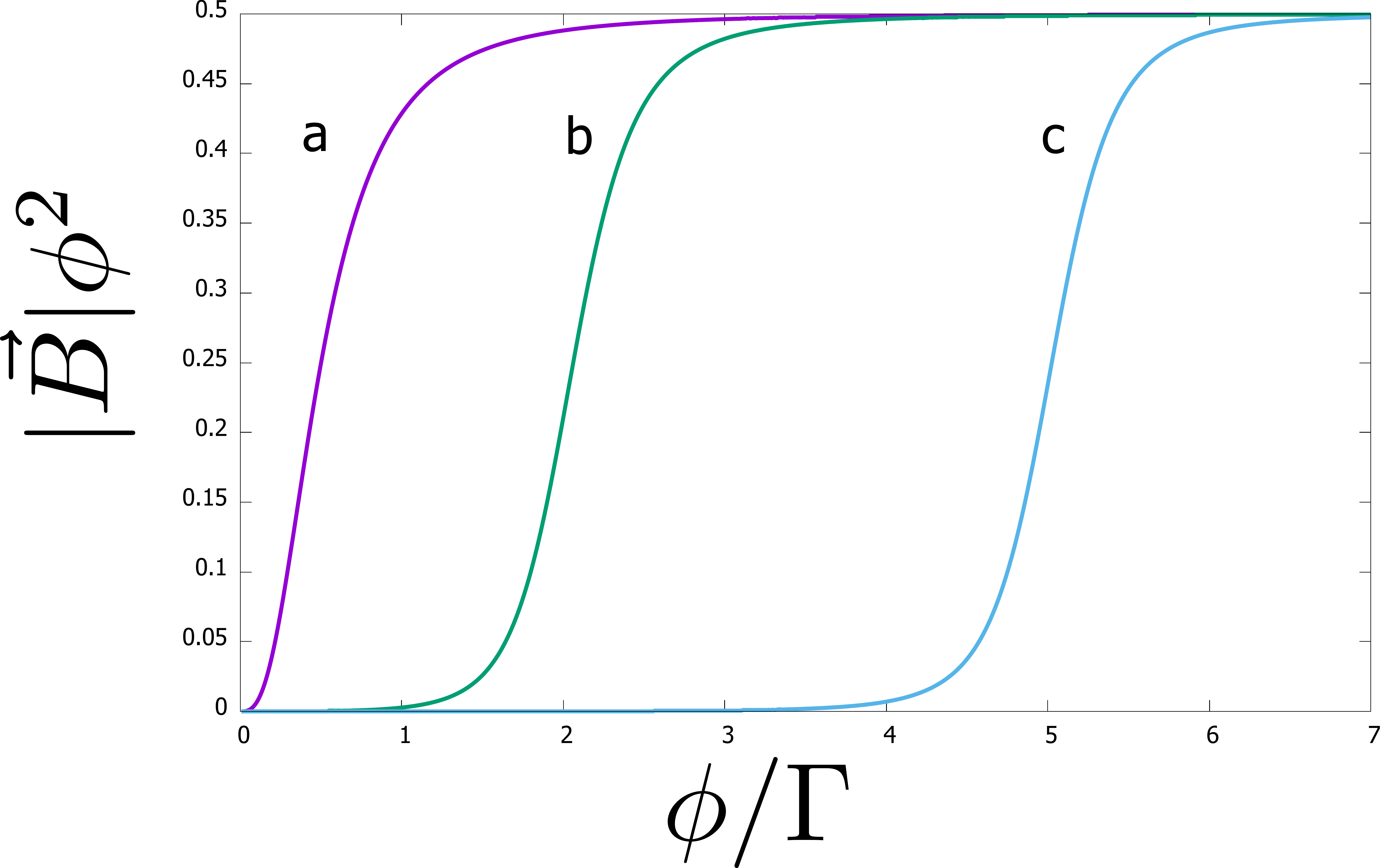}}
\caption{Redefined Berry phase $\times \phi^2$. The curves a,b,c correspond to $B=0,2,5 \Gamma$. They quickly approach the standard expression at $\phi \gg \Gamma$.
}
\end{figure}

Following \cite{Repin2019}, we redefine Berry curvature as an asymmetric part of the response function. For any discrete spectrum and zero temperature, this redefinition would be exact retaining all topological properties of the curvature provided the limit $\omega to 0$ implies $\omega \ll \delta$, $\delta$ being the level spacing in the spectrum. However, in our case the spectrum is continuous, that is, $\delta =0$, and the limit $\omega \to 0$ rather implies $\omega \ll Gamma$. Nevertheless, the redefined curvature coincides with the standard expression at $\phi, B \gg \Gamma$, that is, far from a close vicinity of the point or the resonance $\phi=B$. General expression reads  
\begin{align}
\vec{B} =\frac{\vec{\phi}}{2\pi \phi^3} 
\left[\sum_{\pm}\arctan\frac{2(\phi\pm B)}{\Gamma}+
\frac{\phi^2-\Gamma^2/4 -B^2}{K} \right]
\end{align}
We plot it in Fig. \ref{fig:berry} for several $B$. At the origin, $\vec{B} \propto \vec{\phi}$, the maximum at $B=0$ is $|\vec{B}|\approx 1.2 \Gamma^{-2}$ and is achieved at $\phi \approx 0.3 \Gamma$. 

\begin{figure}
\label{fig:density}
\centerline{\includegraphics[width=0.5\textwidth]{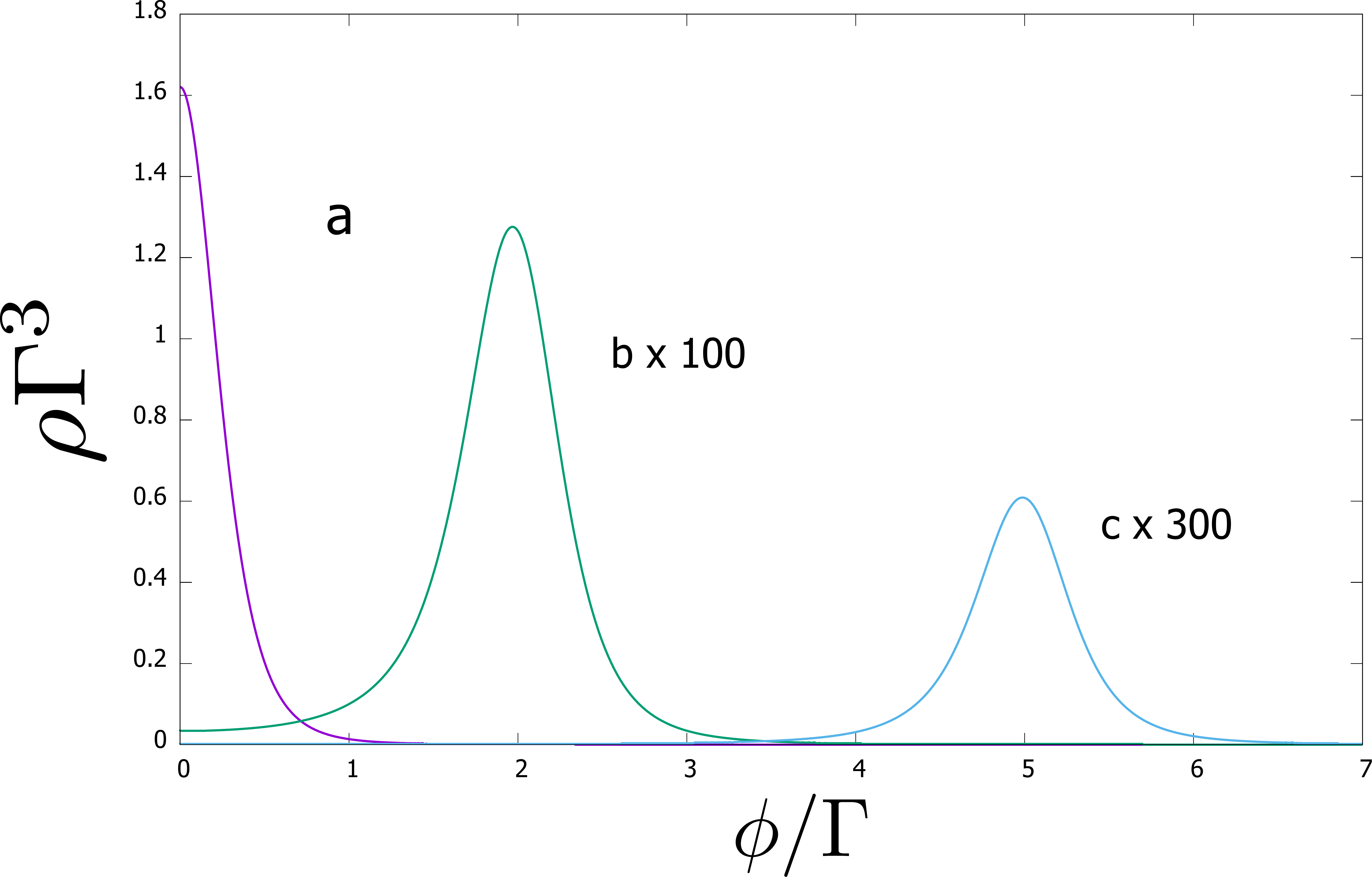}}
\caption{ The density of topological charge. Curves a, b, c correspond to $B=0, 2, 5 \Gamma$, b and c are rescaled as shown in the plot.
}
\end{figure}

So-redefined Berry curvature gives rise to a continuous density of topological charge,
\begin{equation}
\rho(\phi) =\frac{1}{2\pi} {\rm div} \vec{B}
\end{equation}
This is the most important manifestation of embedding a topological singularity into a continous spectrum. The point-line unit charge is spead over the parameter space concentrating either near the origin or, at $B \gg \Gamma$ at the surface $\phi=B$. We evaluate
\begin{equation}
\rho(\phi,B) = \frac{\Gamma^3}{4\pi^2} \frac{B^2 +\phi^2 +\Gamma^2/4}{K^2}
\end{equation}
At small $\Gamma$, the density is proportional to $\Gamma^3$ arising from a complex tunneling process. Its maximum value $\simeq \Gamma^{-3}$ at $B=0$ and $\simeq B^{-2} \Gamma^{-1}$ at $B \gg \Gamma$. 
We plot the density at several values of $B$ in Fig. \ref{fig:density}

\section{Currents in normal leads: pumping}
\label{sec:pumping}
A slow change of control phases may lead to the currents in the normal leads proportional to the time derivatives of the phases, 
\begin{equation}
J_a = e \left(\vec{A}_a(\vec{\phi})\cdot \frac{d{\vec{\phi}}}{d t}\right)
\end{equation}
$\vec{A}_a$ being $\vec{\phi}$-dependent proportionality coefficients. Let us recognize  this as a case of parametric pumping, a phenomenon that has been intensively discussed in quantum transport \cite{Brouwer, QuantumTransport}, also in the context of superconducting nanostructures with normal leads \cite{Blaauboer}. An ac modulation of $\phi$ is expected to result in an ac normal current, that is difficult to measure. However, it can also give rise to a dc current, that is, to pumping. If $\vec{\phi}$ is changing periodically along a closed contour, the charge per cycle depends on the contour only, and, by virtue of Stokes theorem, is given by a flux of the curl of $\vec{A}$ through the contour,   
\begin{equation}
Q_a = \int_0^T dt J_a(t) = \oiint dS (\vec{N}\cdot {\rm curl}\vec{A}).
\end{equation}

We evaluate $\vec{A}$ making use of Eq. \ref{eq:Ja-simple} and expanding the Green functions up to first order in $\check{\delta H}$. We notice that the currents, since the filling factors are in equilibrium, are only due to the vector parts of $\Gamma$. Two groups of terms in Eq. \ref{eq:Ja-simple} that cancel each other in stationary equilibrium case can be rewriten as
\begin{equation}
J_a = \frac{1}{2} {\rm Tr} [ (\vec{\Gamma}_a \cdot \check{\vec{\tau}} [\check{f},\check{G}]\check{\bar{G}}]
\end{equation}
The commutator in this expression in energy representation can be rewritten as 
\begin{equation}
(f(\epsilon)-f(\epsilon-\omega)) \check{G}_{\epsilon, \epsilon -\omega}
\end{equation}
Since we are to expand to the first order in $\omega$, this will give a weight of $\partial_\epsilon f$ in the integration over $\epsilon$, and we can neglect small $\omega$ in the Green functions. The quantities under evaluation just sample Green functions in an energy interval $\simeq k_BT$ near zero energy, this interval going to zero at vanishing temperature. This is in contrast to the response functions explored in the previous Section, those are determined by integration over all relevant energies. Nevertheless, the expression of $\vec{A}$ has qualitatively similar features, the values being concentrated at $\phi \simeq \Gamma$ if $B \ll \Gamma$ or at $\phi=B$  
\begin{align}
\vec{A}_a = -\frac{\Gamma}{\pi K} \left( \vec{\Gamma}_a \Gamma + \vec{\phi}(\vec{\Gamma}_a \cdot \vec{\phi})\Gamma \frac{4 B^2}{K}  + (\Gamma_a \times \vec{\phi}) \right)
\end{align}
Since we discuss the pumping, the curl of $\vec{A}$ --- let us call it the effective field ---is more relevant for us:
\begin{align}
{\rm curl}\vec{A}_a &=& -\frac{\Gamma}{2\pi K^2} [ (\vec{\Gamma}_a \times \vec{\phi}) 4\Gamma(\phi^2 +\Gamma^2/4) \nonumber\\
&+& \vec{\Gamma}_a ((B^2+\Gamma^2/4)^2 - \phi^4) \nonumber\\
&+& \vec{\phi}(\vec{\phi}\cdot \vec{\Gamma}_{a}) (\phi^2 +\Gamma^2 -B^2)].
\end{align}
The natural axis in $\vec{\phi}$ space is set by the direction of $\vec{\Gamma}_a$. In the above expression, we have separated the effective field into azimuthal, axial, and radial component. The dimension of effective field is $E^{-2}$. Far from the resonance, the azimuthal field is estimated as $\simeq \Gamma^3 \phi^{-5}$, and axial/radial field as $\simeq \Gamma^2 \phi^{-4}$. Thus, the typical $Q_a/e$ for the contours that do not cross the resonance are small, $(\Gamma/\phi)^3$, $(\Gamma/phi)^2$ respectively. At the resonance $\phi = B \gg \Gamma$,
the azimuthal field is estimated as $B^{-1}\Gamma^{-1}$, and axial/radial field as $B^{-2}$. At $B \simeq \Gamma$, and near the origin, all field components are estimated as $\Gamma^2$. This implies that we can achieve $Q_a \simeq e$ for small contours with dimension $\Gamma$ provided they are close to the origin.

\begin{figure*}
\label{fig:pumping}
\centerline{\includegraphics[width=0.9\textwidth]{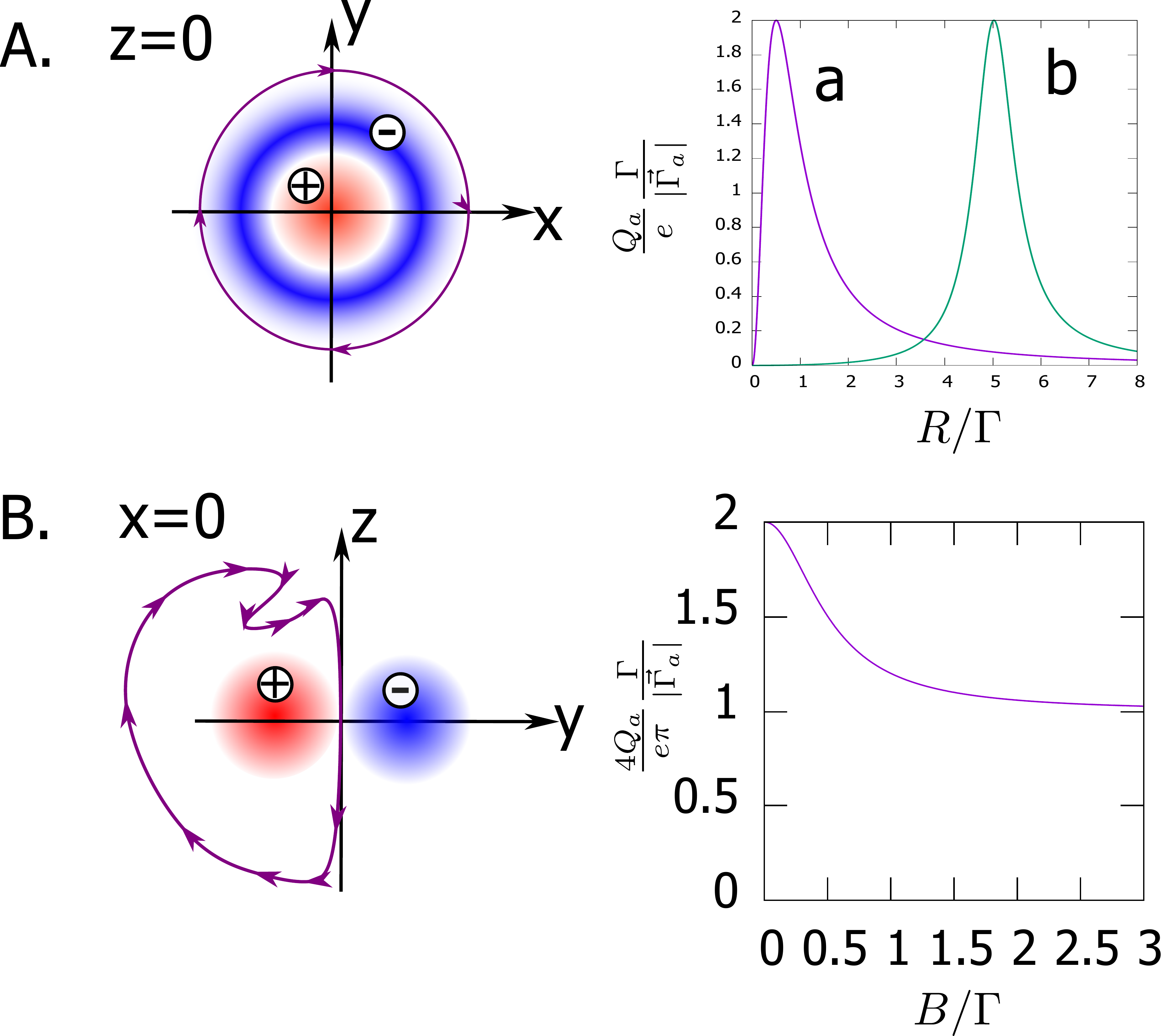}}
\caption{ Pumping to a normal lead, $\vec{\Gamma}_a \parallel z$. A. Probing the axial effective field. A circular contour with radius $R$ in $z=0$ plane is centered at the origin. The plot: dependence of the charge per cycle on $R$ for $B=0,5\Gamma$. B. Probing the azimuthal effective field. The contour in $x=0$ plane that goes along the axis at the scale ${\rm max} B,\Gamma$ encircles the whole flux in this direction. The value of the charge per cycle does not depend on the contour details and is given as function of $B$ in the plot. 
}
\end{figure*}

We illustrate this with the following examples (Fig. \ref{fig:pumping}). 
For pumping in the lead $a$, it is convenient to choose the coordinate system such that $z\parallel \vec{\Gamma}_a$. We probe the axial component of the effective field by taking a circular orbit with radius  $R$ in the plane $z=0$, that is centered at the origin.(Fig. \ref{fig:pumping} a). The axial field is positive at the origin, and changes sign at $\phi=\sqrt{B^2+\Gamma^2/4}$. The total flux in $z=0$ plane is zero. The charge per cycle for this orbit is given by
\begin{equation}
Q_a/e = \frac{2 |\vec{\Gamma}_a| \Gamma R^2}{(R^2+\Gamma^2/4+B^2)^2 - 4R^2B^2}.
\end{equation}
It reaches maximum that does not depend on magnetic field,
\begin{equation}
Q_a = 2e \frac{|\vec{\Gamma}_a|}{\Gamma},
\end{equation}
and gets back to zero for the contours of bigger radius. To probe the azimuthal field, one chooses a contour in e.g. $x=0$ plane, that follows the axis at the scale $\rm{max}(B, \Gamma)$ to enclose the maximum positive flux. The charge per cycle in this case does not depend on the contour details and equals
\begin{equation}
Q_a = - \pi e \frac{|\vec{\Gamma}_a|}{4\Gamma} \frac{\Gamma^2/2 +B^2}{\Gamma^2/4 +B^2}.
\end{equation}
The vector parts of $\Gamma$ are generally different in different leads, so that the same contour is oriented differently for different leads. We conclude that the pumping to the normal leads provides an interesting possibility to explore the vicinity of the Weyl point.

\section{Conclusions}
\label{sec:conclusions}
To conclude, we have investigated the properties of a Weyl point immersed to a continuous spectrum. We take a Weyl point in a superconducting nanostructure that is tunnel-coupled to the electronic states in the normal leads. The tunnel coupling gives rise to a new energy scale $\Gamma$, that corresponds to a scale in parametric space. We investigate in detail how the topological and spectral singularities of the Weyl point are smoothed on this scale. We evaluate the superconducting currents in equilibrium, the superconducting and normal-lead currents at constant voltages applied to the leads. We find sharp features in high-voltage tunnel currents that may be used to detect the Weyl points in experiment.

Importantly, we consider the adiabatic variation of control phases. This permits us to redefine Berry curvature and evaluate the density of topological charge that is not point-like but rather spread around the origin as the manifestation of coupling to the continuous spectrum. 
 
 We investigate the pumping to normal leads and find that it witnesses the peculiarities of Weyl point at the scale of $\Gamma$ and opens up new perspectives for experimental exploration of Weyl point singularities.

\begin{acknowledgements}
This research was supported by the European
Research Council (ERC) under the European Union's
Horizon 2020 research and innovation programme (grant
agreement No. 694272).
\end{acknowledgements}

\bibliographystyle{apsrev4-1}
\bibliography{topi}
\end{document}